# Composite Functional Metasurfaces for Multispectral Achromatic Optics

**Ori Avayu[1#], Euclides Almeida[2#], Yehiam Prior[2*], Tal Ellenbogen[1*]**

Authors' affiliations and contact details:

Ori Avayu[,1],

Tel: +972-3-6408047  Email: oria@post.tau.ac.il

Euclides Almeida[2],

Tel: +972-8-9342245  Email: Euclides.almeida@weizmann.ac.il

Yehiam Prior[2]

Tel: +972-8-9344008  Email: yehiam.prior@weizmann.ac.il

Tal Ellenbogen[*,1],

Tel:  +972-3-6407372  Email: tellenbogen@tauex.tau.ac.il

[1]*Department of Physical Electronics, Fleischman Faculty of Engineering, Tel-Aviv University, Tel-Aviv 69978, Israel*

[2]*Deparement of Chemical Physics, Weizmann Institute of Science, Rehovot, Israel 76100*

#Equal contribution

*Corresponding Authors: tellenbogen@tauex.tau.ac.il , Yehiam.prior@weizmann.ac.il

# Composite Functional Metasurfaces for Multispectral Achromatic Optics

**Ori Avayu[1#], Euclides Almeida[2#], Yehiam Prior[2\*], Tal Ellenbogen[1\*]**

*[1]Department of Physical Electronics, Fleischman Faculty of Engineering, Tel-Aviv University, Tel-Aviv 69978, Israel*

*[2]Department of Chemical Physics, Weizmann Institute of Science, Rehovot, Israel 76100*

*Corresponding authors: tellenbogen@tauex.tau.ac.il, Yehiam.prior@weizmann.ac.il*

**Nanostructured metasurfaces offer unique capabilities for local control of the phase and amplitude of transmitted and reflected optical waves. Based on this potential, a large number of metasurfaces have been proposed in recent years as alternatives to standard optical elements. In most cases, however, these elements suffer from large chromatic aberrations, thus limiting their usefulness for multi-wavelength or broadband applications. Here, in order to alleviate and correct the chromatic aberrations of individual diffractive elements, we introduce dense vertical stacking of independent metasurfaces, where each layer comprises a different material, and is optimally designed for a different band within the visible spectrum. Using this approach, we demonstrate the first triply RGB achromatic metalens in the visible range and perform color imaging with this lens. We further demonstrate functional beam shaping by constructing a self-aligned integrated element for STimulated Emission Depletion (STED) microscopy and a lens that provides anomalous dispersive focusing. These demonstrations lead the way to the realization of superachromatic ultrathin optical elements and multiple functional operations – all in in a single nanostructured ultrathin element.**

Some of the most important technological developments to date rely on the ability of optical elements to control and manipulate the flow of light. While for many years mostly the functionalities of the optical elements, e.g. focusing or beam shaping characteristics, were emphasized, modern technological developments pose strict requirements on the size and thickness of optical elements that are to be used in device applications. Nanotechnology in general and metamaterials or metasurfaces in particular, provide a set of tools that paves the way towards these goals[1–3].

The introduction and development of metasurfaces[2,3] has been one of the most important advances in the design of optical components in recent years. These are ultrathin films, usually a few tens of nanometers thick, composed of dense subwavelength arrays of metallic[4,5] or high index dielectric[2,3,6] resonant scatterers (nanoparticles or nanocavities), that are specifically designed for phase and amplitude control of light. Thus various metasurfaces lenses[6–12], beam shapers[13–15], optical switches[16,17], and even holograms[10,18–27] were demonstrated. Furthermore, non-linear beam generation and beam shaping were studied[28–33], including nonlinear focusing[31,32,34] and holography[35,36].

A lingering problem of metasurface-based optical elements is their strong chromatic aberrations that originate from their diffractive nature. Hence, most of the designs are suitable for only a single wavelength at a specific polarization. Thus, the demonstration of broadband, polarization independent functional elements remains a major challenge. Recently, phase-gradient achromatic metasurfaces have been demonstrated in the near infrared wavelength region[37,38]. The working principle behind those metasurfaces is that each nanoresonator or unit-cell compensates for the phase shifts introduced during light propagation for all desired wavelengths. However, the design of such metasurfaces for arbitrary frequency bands proved to be quite challenging.

Here we propose a conceptually simple yet powerful nanotechnology-driven approach for functional spectral multiplexing of broadband visible light. In our stacked multilayered metasurfaces, each layer is fabricated of different materials and with different design parameters to optimize it for specific frequency band, and if so desired, for a predefined functionality. The layers consist of metallic disc-shaped-nanoparticles that support Localized Surface Plasmon Resonances (LSPR) in the visible part of the spectrum. The dependence of the LSPR on the parameters of the nanodiscs and on their material provides control over the spectral response of the layer so that each one operates independently and with minimal spectral cross talk with the others. We show that multi-layer elements can therefore be designed using simple design rules, and fabricated with readily available nano-lithography processes, thus facilitating the realization of high performance, multifunctional elements. We use this approach to demonstrate an aberration corrected metamaterial-based triplet lens for RGB colors in the visible spectrum, integrated elements for STED microscopy, and elements with anomalous dispersive focusing.

The concept of an aberration corrected multilayer composite structure is illustrated in Fig. 1a. The lens consists of three closely stacked metasurfaces each composed of nanoantennas made of a different metal: gold, silver, and aluminum, and is designed to optimally interact with light at wavelengths of 650nm, 550nm, and 450nm respectively. Each of the layers acts as a narrow band binary Fresnel Zone Plate (FZP) lens that focuses its targeted light to the common focal point. Within each layer the nanoparticles are closely spaced to avoid diffraction-grating effects. For the present lenses, an inter-layer distance of 200 nm was chosen to minimize the near-field cross talk between the individual nanoantennas in the different layers.

The lenses were fabricated by e-beam lithography as described in the Methods section and also in more detail elsewhere[39]. The process involves in-situ consecutive steps of lithography, plasma etching, metal evaporation and plasma-enhanced chemical vapor deposition of silica, which served as the dielectric spacer between the metasurfaces. Similar approaches had been used before to create 3D metasurfaces for other purposes (e.g. [40,41]). The multilayer process allows us to stack the metasurfaces with interlayer stamping precision on the order of tens of nanometers, which is crucial for the performance of some of the functions described below. Figure 1b shows the local design parameters used for the metasurface layers. The respective scanning electron microscope images of the different layers are shown in Supplementary Figure 1. One of the main advances introduced here for the first time to the best of our knowledge, is the use of different metals in different layers in order to optimize the performance of the composite 3D metasurface. Specifically, the use of gold for the red part of the spectrum, silver for the green and aluminum for the blue was a major design advantage in that it allows us to decrease the size of the building blocks in each layer and thus to reduce the spectral cross talk between the different layers. The disk-shape structures were chosen due to their polarization-independent tunable plasmonic resonances. Figure 1c-e show dark-field microscope images of the individual metasurface lenses. The metasurfaces are illuminated with white light and as can be clearly evidenced, each of the metasurfaces strongly scatters light at the designated color.

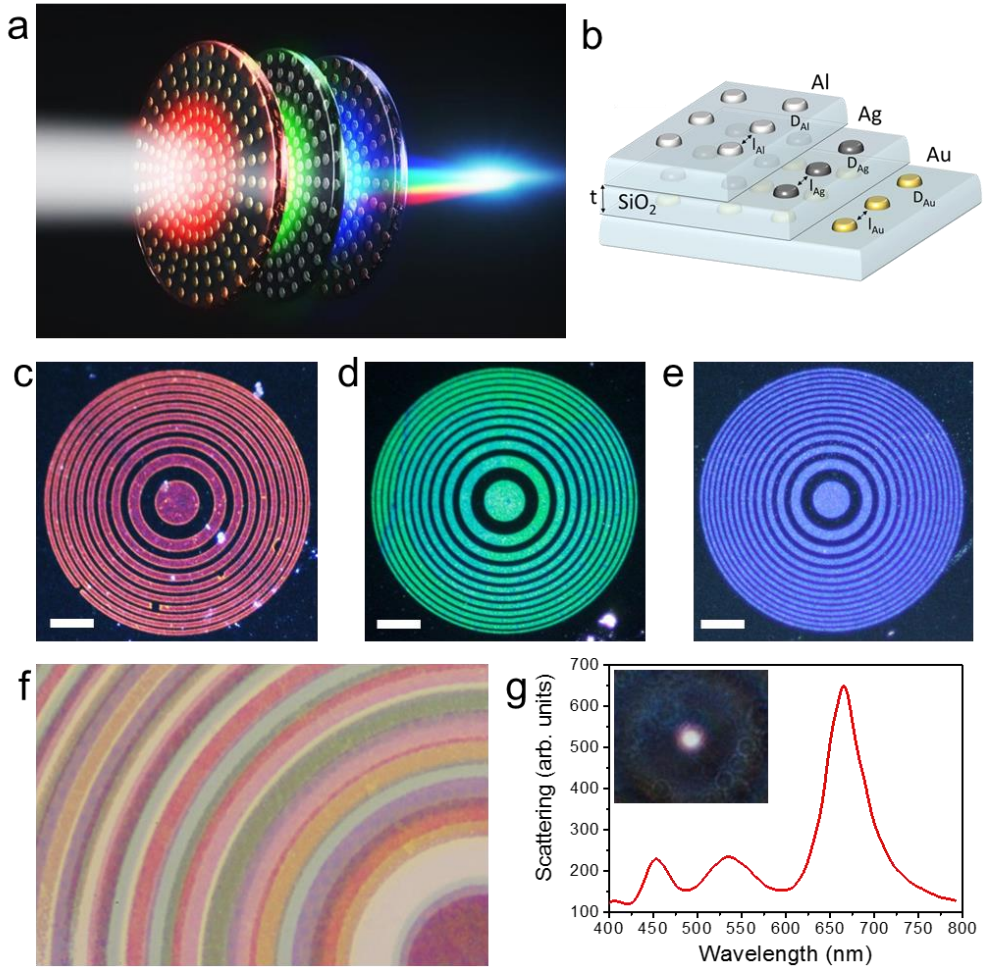

**Figure 1. Three layer lens.** (**a**) Artist's view of the three layer lens. When illuminated with white light, each layer focuses its designated part of the spectrum to a distance of 1 mm along the optical axis. (**b**) Schematic illustration of the layered structure (SEM images of the different layers are given in the Supplementary Material). Each layer consists of nanodiscs with the following diameters D and separations l: $D_{Au}$=125 nm, $l_{Au}$=185 nm; $D_{Ag}$=85 nm, $l_{Ag}$=195 nm; $D_{Al}$=120 nm, $l_{Al}$=150 nm. (**c**) – (**e**) Dark field images of the single-layer lens elements. The different elements are designed to focus red, green or blue to 1mm focal distance along the optical axis. (Scale bars 35 $\mu m$) (**f**) Bright field transmission image of the three layer lens. The rings in each layer block a different color, and the darker areas correspond to regions where the zones overlap. (**g**) A spectrum taken under white light illumination at the focal spot,

revealing the red, green and blue components. The inset shows a colored photograph of the obtained white focal spot at 1 mm along the optical axis.

We used the well-known Fresnel binary zone plate configuration[42] for the individual lenses. The radii of the opaque concentric rings are given by:

$$r_n^2 = n\lambda(f + \frac{n\lambda}{4}) \qquad (1)$$

where $r_n$ is the n[th] zone radius, $\lambda$ is the wavelength and $f$ is the focal distance. Rewriting equation (1) as a function of $f$ gives:

$$f(\lambda) = \left[r_n^2 - \frac{(n\lambda)^2}{4}\right]/(n\lambda) \qquad (2)$$

Equation (2) reveals the dispersive character of the diffractive lenses (see also Supplementary Figure 2). For example, in the case of a conventional zone plate designed to focus green light ($\lambda = 550nm$) to $1mm$ away from the lens, the focal plane of the entire visible spectrum is spanned over more than $400\mu m$. This chromatic aberration is the main hindrance preventing the use of such lenses for broadband or multi-wavelength applications. However, by utilizing frequency selective metasurfaces we can derive a generalized expression for $f(\lambda)$ for each surface:

$$f_i(\lambda) = \left\{\Theta(\lambda - \lambda_{\min,i}) - \Theta(\lambda - \lambda_{\max,i})\right\}f(\lambda) \qquad (3)$$

where $i$ indicates the surface number, $\Theta$ is the Heaviside step function, and $\{\lambda_{min}, \lambda_{max}\}$ is the spectral band of interest. Summing over all surfaces we can obtain the response of the multilayer composite device as follows:

$$f(\lambda) = \sum_i\left\{\Theta(\lambda - \lambda_{\min,i}) - \Theta(\lambda - \lambda_{\max,i})\right\}f_i(\lambda) \qquad (4)$$

In this work we demonstrate this approach by dividing the visible spectrum to 3 spectral bands.

Figure 1f shows a bright-field image of the three-layer element. Note that the zones for each color do not fully overlap, since each layer is designed for a different color to be focused to the same focal position of 1 mm (equation (1)). First we illuminated the lens with white light (Xenon arc lamp), and measured the spectrum at the focal spot. The result is shown in Fig. 1g, where the three designed spectral RGB components at 650nm, 550nm and 450 nm are clearly visible, thus creating the desired white focal spot, as seen in the inset of Fig. 1g. The transmission peak in the red is higher than the other two due to better fabrication and probably also larger wavelength to diameter ratio of the gold nanodiscs. The slight inaccuracies in the target wavelengths are attributed to shifts between the design dimensions and actual fabrication results.

To characterize the focusing properties of the three individual lenses, we used laser illumination at the different wavelengths (see Methods). The results are shown in Fig. 2a, 2b, and 2c for illumination at 450nm, 550nm, and 650 nm respectively. The focal spot diameters for the different colors were measured by sampling 20 points within the expected focal depth region (see Methods and Supplementary Figure 3).  The full width at half maximum (FWHM) for each color at the focal point is measured to be 2.6 $\mu m$, 2.43 $\mu m$ and 2.11 $\mu m$ for red, green and blue wavelengths, respectively (see Figs. 2d and 2e) - in good agreement with theoretical values. The focusing transmission efficiency was also measured for each wavelength, and found to be in the range of  5.8% − 8.7% , which is well within the range of the theoretical value of ∼10% for binary diffractive lenses[42].

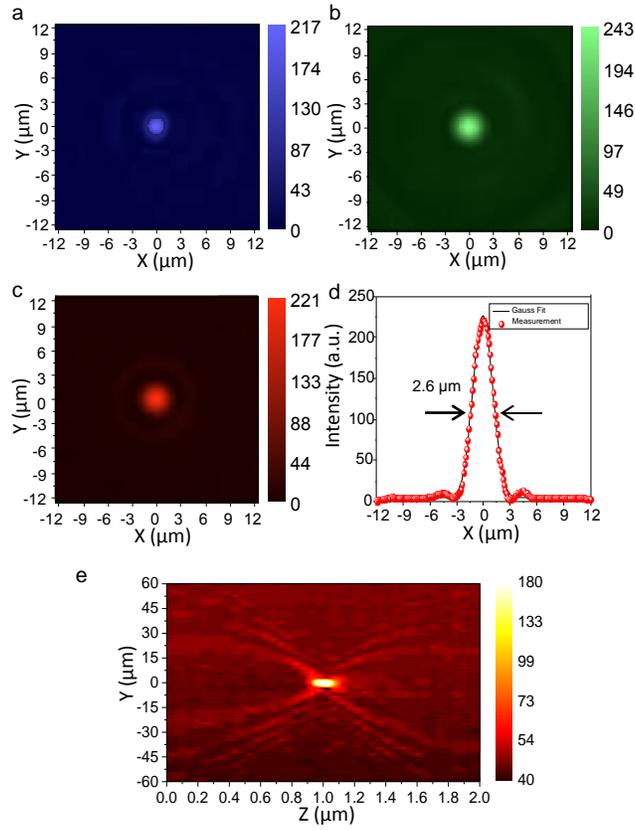

**Figure 2. Focusing lens beam characterization with laser illumination.** (**a**)-(**c**) Focal spots under blue, green, and red laser illuminations. (**d**) Cross section of the beam profile in focus. The data points were fitted to a Gaussian profile. (**e**) Focusing of the lens under green laser illumination.

To compare the broadband operation of the new lens to a conventional binary FZP lens we also fabricated a conventional binary FZP (see Methods), illuminated both lenses with white light (Xenon arc lamp), and characterized the light propagation after the lenses. Figure 3a and 3b show the light propagation after the conventional FZP and the multilayer metasurfaces lens respectively. It can be seen clearly that for the case of the conventional FZP (Fig. 3a) the focus is strongly chromatically aberrated by more than $400\mu m$. For the multilayer metasurfaces lens (Fig. 3b), on the other hand, the chromatic aberrations are corrected and a white focus is formed at 1mm away from the lens. The background of the conventional FZP is darker since it was fabricated as transparent rings in a continuous thin film, thus blocking background illumination. Also its dynamic range is somewhat larger than the fabricated metasurfaces based FZP that show lower extinction compared to continuous films. In Fig. 3c-e we show the performance of conventional FZP with a laser illumination (see Methods) at wavelengths of 450, 550 and 650 nm respectively and compare to the performance of the composite metasurface for the same wavelengths (Fig. 3f-h). The perfect chromatic aberration correction of the composite metasurface at these wavelengths is clear. Figure 3i depicts the measured focal distance vs wavelength for the uncorrected and corrected lenses. This measured low spread of the wavelength-dependent focal plane, and low cross talk between the different layers, enables our lens to perform chromatic imaging, as presented in Fig. 3j (see Methods).

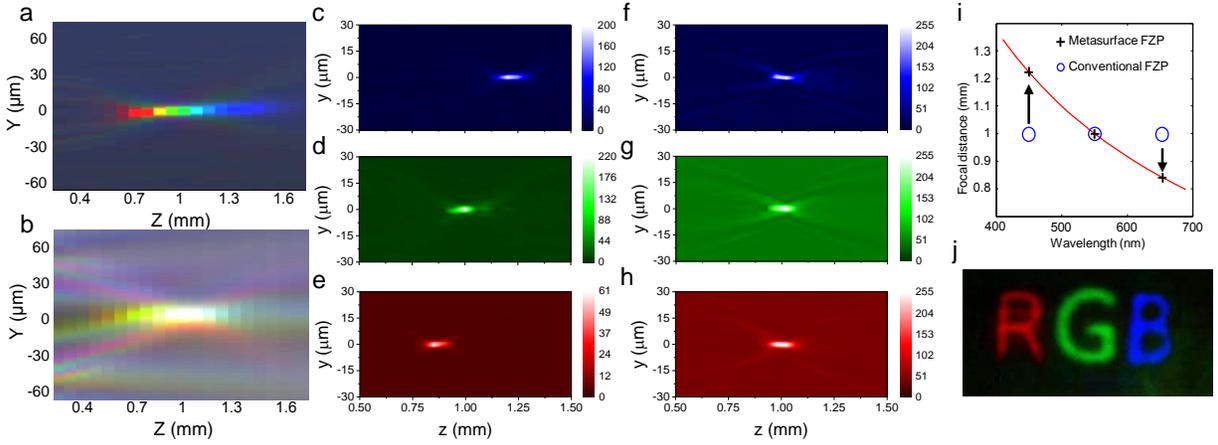

**Figure 3. Chromatically corrected three layer metasurface lens.** Measured light focusing with conventional FZP (**a**) and metasurface FZP (**b**) under white light illumination (Xenon arc lamp, contrast normalized for viewing purposes). Chromatic aberration is apparent in (**a**) while the focal spot at 1 mm appears white in (**b**). Images of the focal region for a conventional FZP illuminated by laser light at 450 nm (c), 550 nm (d) and 650 nm (e) and for the metasureface FZP (f-h), showing the aberration correction for the latter. (i) Theoretical calculation (cf. equation (2)) of the focal distance for a conventional FZP (red line) and the measured focal points at the RGB wavelengths of the conventional FZP (crosses) and metasurface FZP (circles). (j) Demonstration of color imaging using the fabricated metasurface FZP element. See Methods for details.

As we move away from the design wavelengths, the triplet lens shows residual chromatic aberration due to the finite bandwidth of the plasmonic resonators (see Supplementary Fig. 2). At the design wavelengths, however, the residual power of the 'other' wavelengths was measured to be less than 10% of the main beam. This residual cross talk, small as it is, can be further decreased by minimizing fabrication errors or by designing resonators with higher quality factors and therefore sharper linewidths. Here, the advantages of our composite material approach become important. While in principle aluminum or silver nanoresonators can be designed to cover the entire visible range and can be fabricated in a single metasurface, this task would necessitate the use of larger or more complex shapes (such as nanorods) which will be polarization sensitive. These more complex shapes may develop higher-order modes (or multiple resonances) at undesired wavelengths or larger radiative losses that would increase the residual chromatic aberration. The use of different materials for the different spectral regions alleviates these problems.

The multilayered metasurfaces approach allows us also to multiplex several beam manipulation functionalities into a single optical element. To exploit these newly introduced capabilities, we fabricated a multilayer integrated element that can be used for super-resolution STED microscopy[43]. In STED one laser beam with a Gaussian profile is tightly focused to excite a fluorescent sample, and a second, co-aligned doughnut shaped beam with zero intensity at its center depletes the emission by saturating the fluorescent transition. Thus, fluorescence is collected only from the much smaller non-depleted region. Using this method, resolution much better than the diffraction limit had been demonstrated (e.g. references [43,44]). Typically, the optical arrangement of such a system is not trivial, involving an optical setup for generating a doughnut shaped beam, with a fluorescence pump beam at its center, coupled to optics for collection of the fluorescence at the third frequency from the center region.

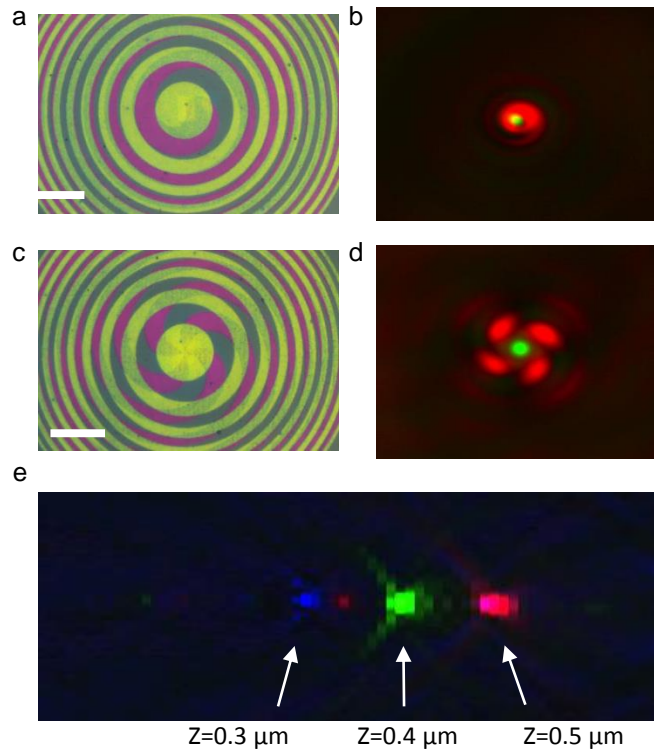

**Figure 4. STED lens** (a) and (b) A dual-layer element composed of one layer designed to focus green laser (λ=550 nm) at $f = 1\ mm$ and a second spiral shaped layer designed to focus a doughnut-shaped red laser beam (λ=650 nm) at the same focal distance. The topological charge of the spiral beam was set to $l = 1$. (**c**) and (**d**) depict the situation for a tightly focused green beam, and a spiral red beam with $l = 4$ both focused to $1\ mm$. The interference between the generated vortex beam and a background beam transmitted by the plate gives rise to the spiral-shaped vortex in (b) and the four lobe vortex in (d). See the recorded propagation of the beams in space showing their red vortex and green focusing characteristics in Supplementary Videos1 and 2. (e) Demonstration of the functionality of a lens that was designed to show anomalous chromatic aberration of its RGB foci, i.e. shorter wavelengths focus before longer wavelengths (Contrast was enhanced for viewing purposes)

Here we utilize the freedom to independently choose different materials and specific designs for each layer, to demonstrate an integrated STED lens, consisting of a dual-layer that tightly focuses green light with a full round beam profile, and red light to a doughnut-shaped beam at the same focal spot. We used a conventional FZP configuration for the excitation focus and fabricated on top a spiral based FZP for the depletion beam that leads to generation of a doughnut beam at its focus[45]. Figure 4a and 4c shows bright field reflection images of two such fabricated devices, where we implement lenses with different topological charges $l = 1$ and $l = 4$ respectively. The performances of the fabricated devices was tested with a super-continuum laser as the illumination source (see Methods), and the results shown in Fig. 4b and 4d, are in good agreement with simulated results.

The ability to design and fabricate a single optical element with different focusing features or different wavelengths, opens the way to a wide range of applications. As another example, we demonstrate a lens with anomalous chromatic dispersion, where the shorter wavelengths are focused first: e.g. blue, green and red colors are focused to 300, 400 and 500 $\mu m$ respectively. The measured performance of such an anomalous lens is shown in Fig. 4e. Such an optical device could, for example, find application in the optical readout of multimedia disks which combine DVD and CD, without the need to use beam splitters or movable lenses. This optical element thus acts as a multiband filter and a lens all-in-one device.

In summary, we have demonstrated a new nanotechnology-driven approach to create thin, multifunctional, and spectrally multiplexed optical elements. We used this approach to demonstrate the first chromatically corrected metasurface triplet lens for RGB colors in the visible, integrated self-aligned STED elements, and a diffractive lens with anomalous chromaticity. These elements are based on a multi-layer design concept, where each layer is composed of nano-resonators made of the dedicated plasmonic material and designed to operate in a specific spectral band. The combined multi-layer elements show complex functionalities, otherwise unachievable with conventional diffractive optics. Due to its simple design, ease of fabrication, ultra-thin profile and low inter-layer crosstalk, our multi-layer, multi-material design could find applications in future integrated opto-electronic devices, imaging systems and complex microscopy setups. Here we have demonstrated the design capabilities in dual- and triple-layer configurations, however this concept can be extended to any number of layers with no additional design complexity. The presented paradigm can be readily implemented also with other materials, building block geometries, and layer designs [6,10,24–26], greatly increasing the composite metasurface efficiency to approach 100%, covering spectral ranges beyond the visible, opening the door to hyperspectral functionality, and addressing specific multi-functional optical requirements.

**Methods:**

*Numerical simulations*

To study the resonant behavior of the metallic nanoantennas, we used a commercially available finite difference time domain simulation software package (Lumerical FDTD). The 3 dimensional simulation was performed with periodic boundary conditions representing a periodic array within each zone of the fabricated optical elements. To simulate the focusing properties of the lenses and STED device, we used a beam propagation algorithm based on the transfer function in free space which was implemented using MATLAB software. See also Supplementary materials.

*Samples fabrication*

The samples were fabricated using multilayer e-beam lithography. We use an Indium-Tin-Oxide (ITO) covered glass as the substrate. A 200 nm thick silica layer was grown on top of the ITO film by plasma-enhanced chemical vapor deposition (PECVD). A 125 nm thick e-beam resist (PMMA 950k A) was then spin coated and the design pattern were exposed in an electron beam lithography. Alignment marks were also written to aid the stamping process of subsequent layers. After development of the resist 30 nm of the exposed silica was etched using inductively coupled plasma.  A 30 nm thick gold film was then deposited by e-beam evaporation and subsequently lifted-off in acetone in an ultrasonic bath. This first layer was covered by a 200 nm thick silica layer grown by PECVD. The process was repeated for the silver and aluminum layers. The aluminum layer was left exposed to the air, where a self-formed passivating aluminum dioxide prevents further degradation of the aluminum disks.

*Experimental setup and measurement of fabricated devices*

We used a Zeiss (Observer Z1) inverted microscope to image the samples in transmission, reflection, bright and dark field modes. To measure the focusing properties we used a home built microscope setup. The emission from the sample was collected with a Mitutoyo 20X 0.42 objective, and the sample was mounted on an automated moving stage (Thorlabs Nanomax 606). We used a Xenon arc lamp as a white light source and a femtosecond optical parametric oscillator (Chameleon OPO VIS, pulse width ~140 fs, repetition rate 80 MHz) as the laser source. We used a similar setup to study the STED element performance, with a super-continuum laser as the illumination source (NKT SuperK compact). The spectral properties of the lenses were obtained using an imaging spectrometer with a cooled back-illuminated EMCCD detector (Andor Shamrock 303i, Newton 970). The imaging of the RGB pattern was done with an LED projector and subsequent optics that were used to project the RGB pattern to the field of view of the fabricated lens.

**Reference list**

1.  Koenderink, A. F., Alu, A. & Polman, A. Nanophotonics: Shrinking light-based technology. *Science (80-. ).* **348,** 516–521 (2015).

2.  Yu, N. & Capasso, F. Flat optics with designer metasurfaces. *Nat. Mater.* **13,** 139–150 (2014).

3.  Kildishev, A. V, Boltasseva, A. & Shalaev, V. M. Planar photonics with metasurfaces. *Science* **339,** 1232009 (2013).

4.  Yu, N. *et al.* Light Propagation with Phase Discontinuities: Generalized Laws of Reflection and Refraction. *Science (80-. ).* **334,** 333–337 (2011).

5.  Ni, X., Emani, N. K., Kildishev, A. V, Boltasseva, A. & Shalaev, V. M. Broadband light bending with plasmonic nanoantennas. *Science* **335,** 427 (2012).

6.  Lin, D., Fan, P., Hasman, E. & Brongersma, M. L. Dielectric gradient metasurface optical elements. *Science (80-. ).* **345,** 298–302 (2014).

7.  Khorasaninejad, M. *et al.* Achromatic Metasurface Lens at Telecommunication Wavelengths. *Nano Lett.* **15,** 5358–62 (2015).

8.  Khorasaninejad, M. *et al.* Metalenses at visible wavelengths: Diffraction-limited focusing and subwavelength resolution imaging. *Science* **352,** 1190–4 (2016).

9.  Chen, X. *et al.* Dual-polarity plasmonic metalens for visible light. *Nat. Commun.* **3,** 1198 (2012).

10. Arbabi, A., Horie, Y., Bagheri, M. & Faraon, A. Dielectric metasurfaces for complete control of phase and polarization with subwavelength spatial resolution and high transmission. *Nat. Nanotechnol.* (2015).

11. Eisenbach, O., Avayu, O., Ditcovski, R. & Ellenbogen, T. Metasurfaces based dual

wavelength diffractive lenses. *Opt. Express* **23,** 3928–36 (2015).

12.    Arbabi, E. *et al.* Multiwavelength metasurfaces through spatial multiplexing. *Sci. Rep.* **6,** 32803 (2016).

13.    Huang, L. *et al.* Dispersionless Phase Discontinuities for Controlling Light Propagation. *Nano Lett.* **12,** 5750–5755 (2012).

14.    Avayu, O., Eisenbach, O., Ditcovski, R. & Ellenbogen, T. Optical metasurfaces for polarization-controlled beam shaping. *Opt. Lett.* **39,** 3892–5 (2014).

15.    Aieta, F. *et al.* Aberration-free ultrathin flat lenses and axicons at telecom wavelengths based on plasmonic metasurfaces. *Nano Lett.* **12,** 4932–6 (2012).

16.    Wen, D. *et al.* Helicity multiplexed broadband metasurface holograms. *Nat. Commun.* **6,** 8241 (2015).

17.    Maguid, E. *et al.* Photonic spin-controlled multifunctional shared-aperture antenna array. *Science* **352,** 1202–6 (2016).

18.    Lin, J., Genevet, P., Kats, M. A., Antoniou, N. & Capasso, F. Nanostructured holograms for broadband manipulation of vector beams. *Nano Lett.* **13,** 4269–74 (2013).

19.    Tsur, Y., Epstein, I. & Arie, A. Arbitrary holographic spectral shaping of plasmonic broadband excitations. *Opt. Lett.* **40,** 1615–1618 (2015).

20.    Walther, B. *et al.* Spatial and Spectral Light Shaping with Metamaterials. *Adv. Mater.* **24,** 6300–6304 (2012).

21.    Larouche, S., Tsai, Y.-J., Tyler, T., Jokerst, N. M. & Smith, D. R. Infrared metamaterial phase holograms. *Nat. Mater.* **11,** 450–454 (2012).

22.    Ni, X. *et al.* Metasurface holograms for visible light. *Nat. Commun.* **4,** 777–778 (2013).

23.    Huang, L. *et al.* Three-dimensional optical holography using a plasmonic metasurface. *Nat.*

*Commun.* **4,** 77–79 (2013).

24.    Zheng, G. *et al.* Metasurface holograms reaching 80% efficiency. *Nat. Nanotechnol.* **10,** 308–312 (2015).

25.    Yifat, Y. *et al.* Highly efficient and broadband wide-angle holography using patch-dipole nanoantenna reflectarrays. *Nano Lett.* **14,** 2485–90 (2014).

26.    Chen, W. T. *et al.* High-efficiency broadband meta-hologram with polarization-controlled dual images. *Nano Lett.* **14,** 225–30 (2014).

27.    Devlin, R. C., Khorasaninejad, M., Chen, W. T., Oh, J. & Capasso, F. Broadband high-efficiency dielectric metasurfaces for the visible spectrum. *Proc. Natl. Acad. Sci.* 201611740 (2016).

28.    Kauranen, M. & Zayats, A. V. Nonlinear plasmonics. *Nat. Photonics* **6,** 737–748 (2012).

29.    Klein, M. W., Enkrich, C., Wegener, M. & Linden, S. Second-harmonic generation from magnetic metamaterials. *Science* **313,** 502–4 (2006).

30.    Salomon, A., Zielinski, M., Kolkowski, R., Zyss, J. & Prior, Y. Size and Shape Resonances in Second Harmonic Generation from Silver Nanocavities. *J. Phys. Chem. C* **117,** 22377–22382 (2013).

31.    Segal, N., Keren-Zur, S., Hendler, N. & Ellenbogen, T. Controlling light with metamaterial-based nonlinear photonic crystals. *Nat. Photonics* **9,** 180–184 (2015).

32.    Almeida, E., Shalem, G. & Prior, Y. Nonlinear Phase Control and Anomalous Phase Matching in Plasmonic Metasurfaces. *Nat. Commun.* (2015).

33.    Wolf, O. *et al.* Phased-array sources based on nonlinear metamaterial nanocavities. *Nat. Commun.* **6,** 7667 (2015).

34.    Tymchenko, M. *et al.* Gradient Nonlinear Pancharatnam-Berry Metasurfaces. *Phys. Rev.*

*Lett.* **115,** 207403 (2015).

35. Ye, W. *et al.* Spin and wavelength multiplexed nonlinear metasurface holography. *Nat. Commun.* **7,** 11930 (2016).

36. Keren-Zur, S., Avayu, O., Michaeli, L. & Ellenbogen, T. Nonlinear Beam Shaping with Plasmonic Metasurfaces. *ACS Photonics* **3,** 117–123 (2016).

37. Aieta, F. *et al.* Multiwavelength achromatic metasurfaces by dispersive phase compensation. *Science (80-. ).* **347,** 1342–1345 (2015).

38. Arbabi, E., Arbabi, A., Kamali, S. M., Horie, Y. & Faraon, A. Multiwavelength polarization-insensitive lenses based on dielectric metasurfaces with meta-molecules. *Optica* **3,** 628 (2016).

39. Almeida, Euclides, Bitton, E., Prior, Y. Nonlinear Metamaterials for Holography. *NCOMMS (in Press. arXiv1512.07899* (2015).

40. Liu, N. *et al.* Three-dimensional photonic metamaterials at optical frequencies. *Nat. Mater.* **7,** 31–7 (2008).

41. Zhao, Y., Belkin, M. A. & Alù, A. Twisted optical metamaterials for planarized ultrathin broadband circular polarizers. *Nat. Commun.* **3,** 870 (2012).

42. Young, M. Zone Plates and Their Aberrations. *JOSA* **62,** 972–976 (1972).

43. Hell, S. W. & Wichmann, J. Breaking the diffraction resolution limit by stimulated emission: stimulated-emission-depletion fluorescence microscopy. *Opt. Lett.* **19,** 780 (1994).

44. Rittweger, E., Han, K. Y., Irvine, S. E., Eggeling, C. & Hell, S. W. STED microscopy reveals crystal colour centres with nanometric resolution. *Nat. Photonics* **3,** 144–147 (2009).

45. Londoño, N., Rueda, E., Gómez, J. A. & Lencina, A. Generation of optical vortices by using

binary vortex producing lenses. *Appl. Opt.* **54,** 796 (2015).

**Acknowledgments:**

This work was supported by the Israeli Ministry of Trade and Labor – Kamin Program, grant.

No. 51387 and was partially supported by the Israeli Science Foundation grants no. 133113 and

124212, by the ICORE program, by an FTA grant from the Israel National Nano Initiative and by the Minerva Foundation.

**Author contributions:**

OA (with help from EA) performed most of the simulations and optical experiments, EA (with help from OA) did most of the design and sample fabrication, all authors discussed the results, analyzed the data and wrote the paper, TE and YP supervised the work.

**Competing interests statement:**

The authors declare that they have no competing financial interests.